\documentclass[letterpaper,compsoc,twoside]{IEEEtran}
\usepackage{fixltx2e} % LaTeX patches, \textsubscript
\usepackage{cmap} % fix search and cut-and-paste in Acrobat
\usepackage{ifthen}
\usepackage[T1]{fontenc}
\usepackage[utf8]{inputenc}
\usepackage{amsmath}

\usepackage[font={small,it},labelfont=bf]{caption}
\usepackage{float}

\usepackage{graphicx}

\usepackage{longtable,ltcaption,array}
\newlength{\DUtablewidth} % internal use in tables
\usepackage{textcomp} % text symbol macros

\usepackage{scipy}
\makeatletter
\def\PY@reset{\let\PY@it=\relax \let\PY@bf=\relax%
    \let\PY@ul=\relax \let\PY@tc=\relax%
    \let\PY@bc=\relax \let\PY@ff=\relax}
\def\PY@tok#1{\csname PY@tok@#1\endcsname}
\def\PY@toks#1+{\ifx\relax#1\empty\else%
    \PY@tok{#1}\expandafter\PY@toks\fi}
\def\PY@do#1{\PY@bc{\PY@tc{\PY@ul{%
    \PY@it{\PY@bf{\PY@ff{#1}}}}}}}
\def\PY#1#2{\PY@reset\PY@toks#1+\relax+\PY@do{#2}}

\def\PY@tok@gd{\def\PY@tc##1{\textcolor[rgb]{0.63,0.00,0.00}{##1}}}
\def\PY@tok@gu{\let\PY@bf=\textbf\def\PY@tc##1{\textcolor[rgb]{0.50,0.00,0.50}{##1}}}
\def\PY@tok@gt{\def\PY@tc##1{\textcolor[rgb]{0.00,0.25,0.82}{##1}}}
\def\PY@tok@gs{\let\PY@bf=\textbf}
\def\PY@tok@gr{\def\PY@tc##1{\textcolor[rgb]{1.00,0.00,0.00}{##1}}}
\def\PY@tok@cm{\let\PY@it=\textit\def\PY@tc##1{\textcolor[rgb]{0.25,0.50,0.56}{##1}}}
\def\PY@tok@vg{\def\PY@tc##1{\textcolor[rgb]{0.73,0.38,0.84}{##1}}}
\def\PY@tok@m{\def\PY@tc##1{\textcolor[rgb]{0.13,0.50,0.31}{##1}}}
\def\PY@tok@mh{\def\PY@tc##1{\textcolor[rgb]{0.13,0.50,0.31}{##1}}}
\def\PY@tok@cs{\def\PY@tc##1{\textcolor[rgb]{0.25,0.50,0.56}{##1}}\def\PY@bc##1{\colorbox[rgb]{1.00,0.94,0.94}{##1}}}
\def\PY@tok@ge{\let\PY@it=\textit}
\def\PY@tok@vc{\def\PY@tc##1{\textcolor[rgb]{0.73,0.38,0.84}{##1}}}
\def\PY@tok@il{\def\PY@tc##1{\textcolor[rgb]{0.13,0.50,0.31}{##1}}}
\def\PY@tok@go{\def\PY@tc##1{\textcolor[rgb]{0.19,0.19,0.19}{##1}}}
\def\PY@tok@cp{\def\PY@tc##1{\textcolor[rgb]{0.00,0.44,0.13}{##1}}}
\def\PY@tok@gi{\def\PY@tc##1{\textcolor[rgb]{0.00,0.63,0.00}{##1}}}
\def\PY@tok@gh{\let\PY@bf=\textbf\def\PY@tc##1{\textcolor[rgb]{0.00,0.00,0.50}{##1}}}
\def\PY@tok@ni{\let\PY@bf=\textbf\def\PY@tc##1{\textcolor[rgb]{0.84,0.33,0.22}{##1}}}
\def\PY@tok@nl{\let\PY@bf=\textbf\def\PY@tc##1{\textcolor[rgb]{0.00,0.13,0.44}{##1}}}
\def\PY@tok@nn{\let\PY@bf=\textbf\def\PY@tc##1{\textcolor[rgb]{0.05,0.52,0.71}{##1}}}
\def\PY@tok@no{\def\PY@tc##1{\textcolor[rgb]{0.38,0.68,0.84}{##1}}}
\def\PY@tok@na{\def\PY@tc##1{\textcolor[rgb]{0.25,0.44,0.63}{##1}}}
\def\PY@tok@nb{\def\PY@tc##1{\textcolor[rgb]{0.00,0.44,0.13}{##1}}}
\def\PY@tok@nc{\let\PY@bf=\textbf\def\PY@tc##1{\textcolor[rgb]{0.05,0.52,0.71}{##1}}}
\def\PY@tok@nd{\let\PY@bf=\textbf\def\PY@tc##1{\textcolor[rgb]{0.33,0.33,0.33}{##1}}}
\def\PY@tok@ne{\def\PY@tc##1{\textcolor[rgb]{0.00,0.44,0.13}{##1}}}
\def\PY@tok@nf{\def\PY@tc##1{\textcolor[rgb]{0.02,0.16,0.49}{##1}}}
\def\PY@tok@si{\let\PY@it=\textit\def\PY@tc##1{\textcolor[rgb]{0.44,0.63,0.82}{##1}}}
\def\PY@tok@s2{\def\PY@tc##1{\textcolor[rgb]{0.25,0.44,0.63}{##1}}}
\def\PY@tok@vi{\def\PY@tc##1{\textcolor[rgb]{0.73,0.38,0.84}{##1}}}
\def\PY@tok@nt{\let\PY@bf=\textbf\def\PY@tc##1{\textcolor[rgb]{0.02,0.16,0.45}{##1}}}
\def\PY@tok@nv{\def\PY@tc##1{\textcolor[rgb]{0.73,0.38,0.84}{##1}}}
\def\PY@tok@s1{\def\PY@tc##1{\textcolor[rgb]{0.25,0.44,0.63}{##1}}}
\def\PY@tok@gp{\let\PY@bf=\textbf\def\PY@tc##1{\textcolor[rgb]{0.78,0.36,0.04}{##1}}}
\def\PY@tok@sh{\def\PY@tc##1{\textcolor[rgb]{0.25,0.44,0.63}{##1}}}
\def\PY@tok@ow{\let\PY@bf=\textbf\def\PY@tc##1{\textcolor[rgb]{0.00,0.44,0.13}{##1}}}
\def\PY@tok@sx{\def\PY@tc##1{\textcolor[rgb]{0.78,0.36,0.04}{##1}}}
\def\PY@tok@bp{\def\PY@tc##1{\textcolor[rgb]{0.00,0.44,0.13}{##1}}}
\def\PY@tok@c1{\let\PY@it=\textit\def\PY@tc##1{\textcolor[rgb]{0.25,0.50,0.56}{##1}}}
\def\PY@tok@kc{\let\PY@bf=\textbf\def\PY@tc##1{\textcolor[rgb]{0.00,0.44,0.13}{##1}}}
\def\PY@tok@c{\let\PY@it=\textit\def\PY@tc##1{\textcolor[rgb]{0.25,0.50,0.56}{##1}}}
\def\PY@tok@mf{\def\PY@tc##1{\textcolor[rgb]{0.13,0.50,0.31}{##1}}}
\def\PY@tok@err{\def\PY@bc##1{\fcolorbox[rgb]{1.00,0.00,0.00}{1,1,1}{##1}}}
\def\PY@tok@kd{\let\PY@bf=\textbf\def\PY@tc##1{\textcolor[rgb]{0.00,0.44,0.13}{##1}}}
\def\PY@tok@ss{\def\PY@tc##1{\textcolor[rgb]{0.32,0.47,0.09}{##1}}}
\def\PY@tok@sr{\def\PY@tc##1{\textcolor[rgb]{0.14,0.33,0.53}{##1}}}
\def\PY@tok@mo{\def\PY@tc##1{\textcolor[rgb]{0.13,0.50,0.31}{##1}}}
\def\PY@tok@mi{\def\PY@tc##1{\textcolor[rgb]{0.13,0.50,0.31}{##1}}}
\def\PY@tok@kn{\let\PY@bf=\textbf\def\PY@tc##1{\textcolor[rgb]{0.00,0.44,0.13}{##1}}}
\def\PY@tok@o{\def\PY@tc##1{\textcolor[rgb]{0.40,0.40,0.40}{##1}}}
\def\PY@tok@kr{\let\PY@bf=\textbf\def\PY@tc##1{\textcolor[rgb]{0.00,0.44,0.13}{##1}}}
\def\PY@tok@s{\def\PY@tc##1{\textcolor[rgb]{0.25,0.44,0.63}{##1}}}
\def\PY@tok@kp{\def\PY@tc##1{\textcolor[rgb]{0.00,0.44,0.13}{##1}}}
\def\PY@tok@w{\def\PY@tc##1{\textcolor[rgb]{0.73,0.73,0.73}{##1}}}
\def\PY@tok@kt{\def\PY@tc##1{\textcolor[rgb]{0.56,0.13,0.00}{##1}}}
\def\PY@tok@sc{\def\PY@tc##1{\textcolor[rgb]{0.25,0.44,0.63}{##1}}}
\def\PY@tok@sb{\def\PY@tc##1{\textcolor[rgb]{0.25,0.44,0.63}{##1}}}
\def\PY@tok@k{\let\PY@bf=\textbf\def\PY@tc##1{\textcolor[rgb]{0.00,0.44,0.13}{##1}}}
\def\PY@tok@se{\let\PY@bf=\textbf\def\PY@tc##1{\textcolor[rgb]{0.25,0.44,0.63}{##1}}}
\def\PY@tok@sd{\let\PY@it=\textit\def\PY@tc##1{\textcolor[rgb]{0.25,0.44,0.63}{##1}}}

\def\PYZus{\char`\_}

\def\PYZsh{\char`\#}

\makeatother

\providecommand*{\DUfootnotemark}[3]{%
  \raisebox{1em}{\hypertarget{#1}{}}%
  \hyperlink{#2}{\textsuperscript{#3}}%
}
\providecommand{\DUfootnotetext}[4]{%
  \begingroup%
  \renewcommand{\thefootnote}{%
    \protect\raisebox{1em}{\protect\hypertarget{#1}{}}%
    \protect\hyperlink{#2}{#3}}%
  \footnotetext{#4}%
  \endgroup%
}

\providecommand*{\DUrole}[2]{%
  \ifcsname DUrole#1\endcsname%
    \csname DUrole#1\endcsname{#2}%
  \else% backwards compatibility: try \docutilsrole#1{#2}
    \ifcsname docutilsrole#1\endcsname%
      \csname docutilsrole#1\endcsname{#2}%
    \else%
      #2%
    \fi%
  \fi%
}

\providecommand*{\DUroletitlereference}[1]{\textsl{#1}}

\ifthenelse{\isundefined{\hypersetup}}{
  \usepackage[colorlinks=true,linkcolor=blue,urlcolor=blue]{hyperref}
  \urlstyle{same} % normal text font (alternatives: tt, rm, sf)
}{}

\begin{document}
\newcounter{footnotecounter}\title{cphVB: A System for Automated Runtime Optimization and Parallelization of Vectorized Applications}\author{Mads Ruben Burgdorff Kristensen$^{\setcounter{footnotecounter}{1}\fnsymbol{footnotecounter}\setcounter{footnotecounter}{2}\fnsymbol{footnotecounter}}$%
          \setcounter{footnotecounter}{1}\thanks{\fnsymbol{footnotecounter} %
          Corresponding author: \protect\href{mailto:madsbk@nbi.dk}{madsbk@nbi.dk}}\setcounter{footnotecounter}{2}\thanks{\fnsymbol{footnotecounter} University of Copenhagen}, Simon Andreas Frimann Lund$^{\setcounter{footnotecounter}{2}\fnsymbol{footnotecounter}}$, Troels Blum$^{\setcounter{footnotecounter}{2}\fnsymbol{footnotecounter}}$, Brian Vinter$^{\setcounter{footnotecounter}{2}\fnsymbol{footnotecounter}}$\thanks{%

          \noindent%
          Copyright\,\copyright\,2012 Mads Ruben Burgdorff Kristensen et al. This is an open-access article distributed under the terms of the Creative Commons Attribution License, which permits unrestricted use, distribution, and reproduction in any medium, provided the original author and source are credited.%
        }}\maketitle
          \renewcommand{\leftmark}{PROC. OF THE 11th PYTHON IN SCIENCE CONF. (SCIPY 2012)}
          \renewcommand{\rightmark}{CPHVB: A SYSTEM FOR AUTOMATED RUNTIME OPTIMIZATION AND PARALLELIZATION OF VECTORIZED APPLICATIONS}

\InputIfFileExists{page_numbers.tex}{}{}
\newcommand*{\docutilsroleref}{\ref}
\newcommand*{\docutilsrolelabel}{\label}
\begin{abstract}Modern processor architectures, in addition to having still more cores, also require still more consideration to memory-layout in order to run at full capacity.
The usefulness of most languages is deprecating as their abstractions, structures or objects are hard to map onto modern processor architectures efficiently.

The work in this paper introduces a new abstract machine framework, cphVB, that enables vector oriented high-level programming languages to map onto a broad range of architectures efficiently. The idea is to close the gap between high-level languages and hardware optimized low-level implementations. By translating high-level vector operations into an intermediate vector bytecode, cphVB enables specialized vector engines to efficiently execute the vector operations.

The primary success parameters are to maintain a complete abstraction from low-level details and to provide efficient code execution across different, modern, processors. We evaluate the presented design through a setup that targets multi-core CPU architectures. We evaluate the performance of the implementation using Python implementations of well-known algorithms: a jacobi solver, a kNN search, a shallow water simulation and a synthetic stencil simulation. All demonstrate good performance.\end{abstract}

%___________________________________________________________________________

\subsection*{\phantomsection%
  Introduction%
  \addcontentsline{toc}{subsection}{Introduction}%
  \label{introduction}%
}

Obtaining high performance from today's computing environments requires both a deep and broad working knowledge on computer architecture, communication paradigms and programming interfaces. Today's computing environments are highly heterogeneous consisting of a mixture of CPUs, GPUs, FPGAs and DSPs orchestrated in a wealth of architectures and lastly connected in numerous ways.

Utilizing this broad range of architectures manually requires programming specialists and is a very time-consuming task – time and specialization a scientific researcher typically does not have. A high-productivity language that allows rapid prototyping and still enables efficient utilization of a broad range of architectures is clearly preferable.
There exist high-productivity language and libraries that automatically utilize parallel architectures \cite{Kri10,Dav04,New11}. They are however still few in numbers and have one problem in common. They are closely coupled to both the front-end, i.e. programming language and IDE, and the back-end, i.e. computing device, which makes them interesting only to the few using the exact combination of front and back-end.

A tight coupling between front-end technology and back-end presents another problem; the usefulness of the developed program expires as soon as the back-end does. With the rapid development of hardware architectures the time spend on implementing optimized programs for specific hardware, is lost as soon as the hardware product expires.

In this paper, we present a novel approach to the problem of closing the gap between high-productivity languages and parallel architectures, which allows a high degree of modularity and reusability. The approach involves creating a framework, cphVB\DUfootnotemark{id4}{id5}{*} (Copenhagen Vector Bytecode). cphVB defines a clear and easy to understand intermediate bytecode language and provides a runtime environment for executing the bytecode. cphVB also contains a protocol to govern the safe, and efficient exchange, creation, and destruction of model data.

cphVB provides a retargetable framework in which the user can write programs utilizing whichever cphVB supported programming interface they prefer and run the program on their own workstation while doing prototyping, such as testing correctness and functionality of their programs. Users can then deploy exactly the same program in a more powerful execution environment without changing a single line of code and thus effectively solve greater problem sets.

The rest of the paper is organized as follows. In Section \DUroletitlereference{Programming Model}. we describe the programming model supported by cphVB. The section following gives a brief description of \DUroletitlereference{Numerical Python}, which is the first programming interface that fully supports cphVB. Sections \DUroletitlereference{Design} and \DUroletitlereference{Implementation} cover the overall cphVB design and an implementation of it. In Section \DUroletitlereference{Performance Study}, we conduct an evaluation of the implementation. Finally, in Section \DUroletitlereference{Future Work} and \DUroletitlereference{Conclusion} we discuss future work and conclude.%
\DUfootnotetext{id5}{id4}{*}{
Open Source Project - Website: \url{http://cphvb.bitbucket.org}.}

%___________________________________________________________________________

\subsubsection*{\phantomsection%
  Related Work%
  \addcontentsline{toc}{subsubsection}{Related Work}%
  \label{related-work}%
}

The key motivation for cphVB is to provide a framework for the utilization of heterogeneous computing systems with the goal of obtaining high-performance, high-productivity and high-portability ($HP^3$). Systems such as pyOpenCL/pyCUDA \cite{Klo09} provides a direct mapping from front-end language to the optimization target. In this case, providing the user with direct access to the low-level systems OpenCL \cite{Khr10} and CUDA \cite{Nvi10} from the high-level language Python \cite{Ros10}.
The work in \cite{Klo09} enables the user to write a low-level implementation in a high-productivity language. The goal is similar to cphVB – the approach however is entirely different. cphVB provides a means to hide low-level target specific code behind a programming model and providing a framework and runtime environment to support it.

Intel Math Kernel Library \cite{Int08} is in this regard more comparable to cphVB. Intel MKL is a programming library providing utilization of multiple targets ranging from a single-core CPU to a multi-core shared memory CPU and even to a cluster of computers all through the same programming API. However, cphVB is not only a programming library it is a runtime system providing support for a vector oriented programming model. The programming model is well-known from high-productivity languages such as MATLAB \cite{Mat10}, \cite{Rrr11}, \cite{Idl00}, GNU Octave \cite{Oct97} and Numerical Python (NumPy) \cite{Oli07} to name a few.

cphVB is more closely related to the work described in \cite{Gar10}, here a compilation framework is provided for execution in a hybrid environment consisting of both CPUs and GPUs. Their framework uses a Python/NumPy based front-end that uses Python decorators as hints to do selective optimizations. cphVB similarly provides a NumPy based front-end and equivalently does selective optimizations.
However, cphVB uses a slightly less obtrusive approach; program selection hints are sent from the front-end via the NumPy-bridge. This approach provides the advantage that any existing NumPy program can run unaltered and take advantage of cphVB without changing a single line of code. Whereas unPython requires the user to manually modify the source code by applying hints in a manner similar to that of OpenMP \cite{Pas05}. This non-obtrusive design at the source level is to the author's knowledge novel.

Microsoft Accelerator \cite{Dav04} introduces ParallelArray, which is similar to the utilization of the NumPy arrays in cphVB but there are strict limitations to the utilization of ParallelArrays. ParallelArrays does not allow the use of direct indexing, which means that the user must copy a ParallelArray into a conventional array before indexing. cphVB instead allows indexed operations and additionally supports \textbf{array-views}, which are array-aliases that provide multiple ways to access the same chunk of allocated memory. Thus, the data structure in cphVB is highly flexible and provides elegant programming solutions for a broad range of numerical algorithms.
Intel provides a similar approach called Intel Array Building Blocks (ArBB) \cite{New11} that provides retargetability and dynamic compilation. It is thereby possible to utilize heterogeneous architectures from within standard C++.
The retargetability aspect of Intel ArBB is represented in cphVB as a plain and simple configuration file that define the cphVB runtime environment. Intel ArBB provides a high performance library that utilizes a heterogeneous environment and hides the low-level details behind a vector oriented programming model similar to cphVB. However, ArBB only provides access to the programming model via C++ whereas cphVB is not biased towards any one specific front-end language.

On multiple points cphVB is closely related in functionality and goals to the SEJITS \cite{Cat09} project. SEJITS takes a different approach towards the front-end and programming model. SEJITS provides a rich set of computational kernels in a high-productivity language such as Python or Ruby. These kernels are then specialized towards an optimality criteria. This approach has shown to provide performance that at times out-performs even hand-written specialized code towards a given architecture. Being able to construct computational kernels is a core issue in data-parallel programming.

The programming model in cphVB does not provide this kernel methodology. cphVB has a strong NumPy heritage which also shows in the programming model. The advantage is easy adaptability of the cphVB programming model for users of NumPy, Matlab, Octave and R. The cphVB programming model is not a stranger to computational kernels – cphVB deduce computational kernels at runtime by inspecting the vector bytecode generated by the Bridge.

cphVB provides in this sense a virtual machine optimized for execution of vector operations, previous work \cite{And08} was based on a complete virtual machine for generic execution whereas cphVB provides an optimized subset.

%___________________________________________________________________________

\subsection*{\phantomsection%
  Numerical Python%
  \addcontentsline{toc}{subsection}{Numerical Python}%
  \label{numerical-python}%
}

Before describing the design of cphVB, we will briefly go through Numerical Python (NumPy) \cite{Oli07}. Numerical Python heavily influenced many design decisions in cphVB – it also uses a vector oriented programming model as cphVB.

NumPy is a library for numerical operations in Python, which is implemented in the C programming language. NumPy provides the programmer with a multidimensional array object and a whole range of supported array operations. By using the array operations, NumPy takes advantage of efficient C-implementations while retaining the high abstraction level of Python.

NumPy uses an array syntax that is based on the Python list syntax. The arrays are indexed positionally, 0 through length – 1, where negative indexes is used for indexing in the reversed order. Like the list syntax in Python, it is possible to index multiple elements. All indexing that represents more than one element returns a view of the elements rather than a new copy of the elements. It is this view semantic that makes it possible to implement a stencil operation as illustrated in Figure \DUrole{ref}{fig-stencil-expr} and demonstrated in the code example below. In order to force a real array copy rather than a new array reference NumPy provides the ''copy'' method.

In the rest of this paper, we define the \textbf{array-base} as the originally allocated array that lies contiguously in memory. In addition, we will define the \textbf{array-view} as a view of the elements in an \textbf{array-base}. An \textbf{array-view} is usually a subset of the elements in the \textbf{array-base} or a re-ordering such as the reverse order of the elements or a combination.\begin{figure}[]
\noindent{\includegraphics[width=\columnwidth]{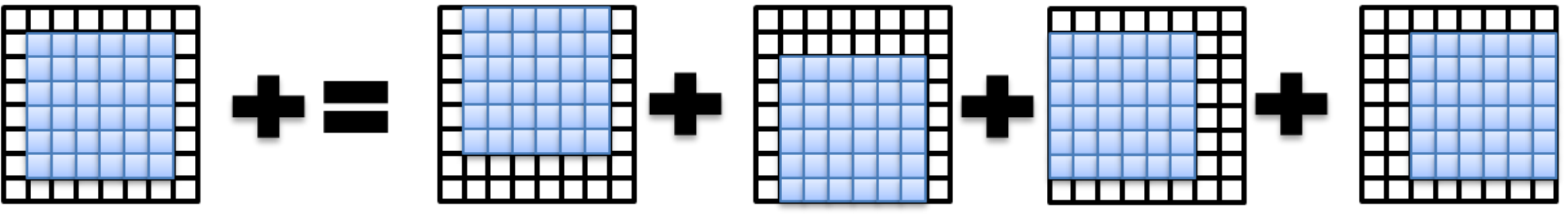}\hfill}
\caption{Matrix expression of a simple 5-point stencil computation example. See line eight in the code example, for the Python expression. \DUrole{label}{fig-stencil-expr}}
\end{figure}
\begin{Verbatim}[commandchars=\\\{\},numbers=left,firstnumber=1,stepnumber=1,fontsize=\footnotesize,xleftmargin=2.25mm,numbersep=3pt]
\PY{n}{center} \PY{o}{=} \PY{n}{full}\PY{p}{[}\PY{l+m+mi}{1}\PY{p}{:}\PY{o}{-}\PY{l+m+mi}{1}\PY{p}{,} \PY{l+m+mi}{1}\PY{p}{:}\PY{o}{-}\PY{l+m+mi}{1}\PY{p}{]}
\PY{n}{up}     \PY{o}{=} \PY{n}{full}\PY{p}{[}\PY{l+m+mi}{0}\PY{p}{:}\PY{o}{-}\PY{l+m+mi}{2}\PY{p}{,} \PY{l+m+mi}{1}\PY{p}{:}\PY{o}{-}\PY{l+m+mi}{1}\PY{p}{]}
\PY{n}{down}   \PY{o}{=} \PY{n}{full}\PY{p}{[}\PY{l+m+mi}{2}\PY{p}{:}  \PY{p}{,} \PY{l+m+mi}{1}\PY{p}{:}\PY{o}{-}\PY{l+m+mi}{1}\PY{p}{]}
\PY{n}{left}   \PY{o}{=} \PY{n}{full}\PY{p}{[}\PY{l+m+mi}{1}\PY{p}{:}\PY{o}{-}\PY{l+m+mi}{1}\PY{p}{,} \PY{l+m+mi}{0}\PY{p}{:}\PY{o}{-}\PY{l+m+mi}{2}\PY{p}{]}
\PY{n}{right}  \PY{o}{=} \PY{n}{full}\PY{p}{[}\PY{l+m+mi}{1}\PY{p}{:}\PY{o}{-}\PY{l+m+mi}{1}\PY{p}{,} \PY{l+m+mi}{2}\PY{p}{:}  \PY{p}{]}
\PY{k}{while} \PY{n}{epsilon} \PY{o}{<} \PY{n}{delta}\PY{p}{:}
    \PY{n}{work}\PY{p}{[}\PY{p}{:}\PY{p}{]} \PY{o}{=} \PY{n}{center}
    \PY{n}{work} \PY{o}{+}\PY{o}{=} \PY{l+m+mf}{0.2} \PY{o}{*} \PY{p}{(}\PY{n}{up}\PY{o}{+}\PY{n}{down}\PY{o}{+}\PY{n}{left}\PY{o}{+}\PY{n}{right}\PY{p}{)}
    \PY{n}{center}\PY{p}{[}\PY{p}{:}\PY{p}{]} \PY{o}{=} \PY{n}{work}
\end{Verbatim}

%___________________________________________________________________________

\subsection*{\phantomsection%
  Target Programming Model%
  \addcontentsline{toc}{subsection}{Target Programming Model}%
  \label{target-programming-model}%
}
To hide the complexities of obtaining high-performance from a heterogeneous environment any given system must provide a meaningful high-level abstraction. This can be realized in the form of domain specific languages, embedded languages, language extensions, libraries, APIs etc. Such an abstraction serves two purposes: 1) It must provide meaning for the end-user such that the goal of high-productivity can be met with satisfaction. 2) It must provide an abstraction that consists of a sufficient amount of information for the system to optimize its utilization.

cphVB is not biased towards any specific choice of abstraction or front-end technology as long as it is compatible with a vector oriented programming model. This provides means to use cphVB in functional programming languages, provide a front-end with a strict mathematic notation such as APL \cite{Apl00} or a more relaxed syntax such as MATLAB.

The vector oriented programming model encourages expressing programs in the form of high-level array operations, e.g. by expressing the addition of two arrays using one high-level function instead of computing each element individually. The NumPy application in the code example above figure \DUrole{ref}{fig-stencil-expr} is a good example of using the vector oriented programming model.

%___________________________________________________________________________

\subsection*{\phantomsection%
  Design of cphVB%
  \addcontentsline{toc}{subsection}{Design of cphVB}%
  \label{design-of-cphvb}%
}

The key contribution in this paper is a framework, cphVB, that support a vector oriented programming model. The idea of cphVB is to provide the mechanics to seamlessly couple a programming language or library with an architecture-specific implementation of vectorized operations.

cphVB consists of a number of components that communicate using a simple protocol. Components are allowed to be architecture-specific but they are all interchangeable since all uses the same communication protocol. The idea is to make it possible to combine components in a setup that perfectly match a specific execution environment. cphVB consist of the following components:%
\begin{description}
\item[{Programming Interface}] \leavevmode 

The programming language or library exposed to the user. cphVB was initially meant as a computational back-end for the Python library NumPy, but we have generalized cphVB to potential support all kinds of languages and libraries. Still, cphVB has design decisions that are influenced by NumPy and its representation of vectors/matrices.
\item[{Bridge}] \leavevmode 

The role of the Bridge is to integrate cphVB into existing languages and libraries. The Bridge generates the cphVB bytecode that corresponds to the user-code.
\item[{Vector Engine}] \leavevmode 

The Vector Engine is the architecture-specific implementation that executes cphVB bytecode.
\item[{Vector Engine Manager}] \leavevmode 

The Vector Engine Manager manages data location and ownership of vectors. It also manages the distribution of computing jobs between potentially several Vector Engines, hence the name.
\end{description}

An overview of the design can be seen in Figure \DUrole{ref}{fig-cphvb-design}.\begin{figure}[]
\noindent{\includegraphics[width=\columnwidth]{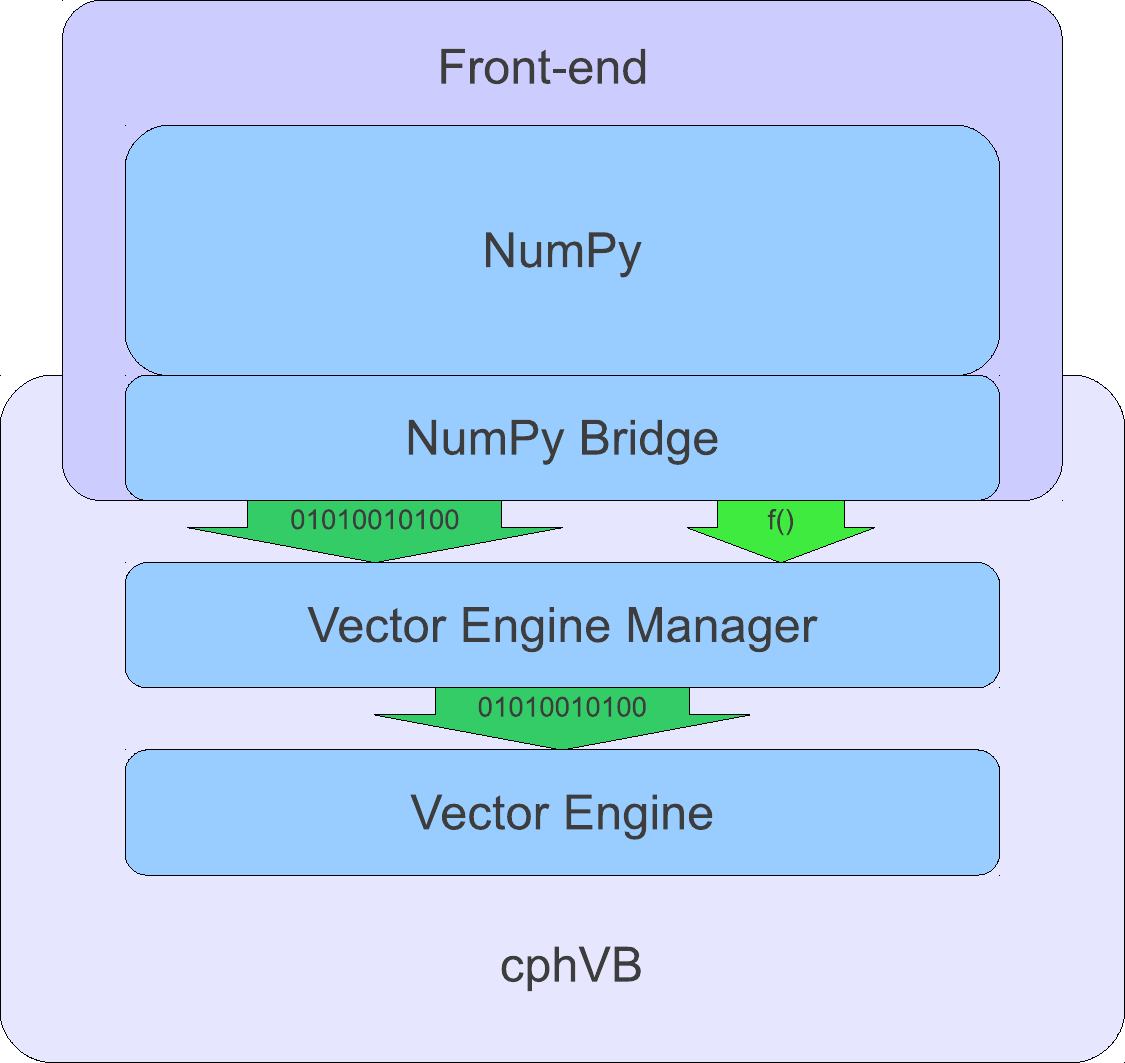}\hfill}
\caption{cphVB design idea. \DUrole{label}{fig-cphvb-design}}
\end{figure}

%___________________________________________________________________________

\subsubsection*{\phantomsection%
  Configuration%
  \addcontentsline{toc}{subsubsection}{Configuration}%
  \label{configuration}%
}

To make cphVB as flexible a framework as possible, we manage the setup of all the components at runtime through a configuration file. The idea is that the user can change the setup of components simply by editing the configuration file before executing the user application. Additionally, the user only has to change the configuration file in order to run the application on different systems with different computational resources. The configuration file uses the ini syntax, an example is provided below.\begin{Verbatim}[commandchars=\\\{\},fontsize=\footnotesize]
\PY{c+cp}{\PYZsh{}}\PY{c+cp}{ Root of the setup}
\PY{p}{[}\PY{n}{setup}\PY{p}{]}
\PY{n}{bridge} \PY{o}{=} \PY{n}{numpy}
\PY{n}{debug} \PY{o}{=} \PY{n+nb}{true}

\PY{c+cp}{\PYZsh{}}\PY{c+cp}{ Bridge for NumPy}
\PY{p}{[}\PY{n}{numpy}\PY{p}{]}
\PY{n}{type} \PY{o}{=} \PY{n}{bridge}
\PY{n}{children} \PY{o}{=} \PY{n}{node}

\PY{c+cp}{\PYZsh{}}\PY{c+cp}{ Vector Engine Manager for a single machine}
\PY{p}{[}\PY{n}{node}\PY{p}{]}
\PY{n}{type} \PY{o}{=} \PY{n}{vem}
\PY{n}{impl} \PY{o}{=} \PY{n}{libcphvb\PYZus{}vem\PYZus{}node}\PY{p}{.}\PY{n}{so}
\PY{n}{children} \PY{o}{=} \PY{n}{mcore}

\PY{c+cp}{\PYZsh{}}\PY{c+cp}{ Vector Engine using TLP on shared memory}
\PY{p}{[}\PY{n}{mcore}\PY{p}{]}
\PY{n}{type} \PY{o}{=} \PY{n}{ve}
\PY{n}{impl} \PY{o}{=} \PY{n}{libcphvb\PYZus{}ve\PYZus{}mcore}\PY{p}{.}\PY{n}{so}
\end{Verbatim}
This example configuration provides a setup for utilizing a shared memory machine with thread-level-parallelism (TLP) on one machine by instructing the vector engine manager to use a single multi-core TLP engine.

%___________________________________________________________________________

\subsubsection*{\phantomsection%
  Bytecode%
  \addcontentsline{toc}{subsubsection}{Bytecode}%
  \label{bytecode}%
}

The central part of the communication between all the components in cphVB is vector bytecode. The goal with the bytecode language is to be able to express operations on multidimensional vectors. Taking inspiration from single instruction, multiple data (SIMD) instructions but adding structure to the data. This, of course, fits very well with the array operations in NumPy but is not bound nor limited to these.

We would like the bytecode to be a concept that is easy to explain and understand. It should have a simple design that is easy to implement. It should be easy and inexpensive to generate and decode. To fulfill these goals we chose a design that conceptually is an assembly language where the operands are multidimensional vectors. Furthermore, to simplify the design the assembly language should have a one-to-one mapping between instruction mnemonics and opcodes.

In the basic form, the bytecode instructions are primitive operations on data, e.g. addition, subtraction, multiplication, division, square root etc. As an example, let us look at addition. Conceptually it has the form:%
\begin{quote}\begin{verbatim}
add $d, $a, $b
\end{verbatim}

\end{quote}
Where \texttt{add} is the opcode for addition. After execution \texttt{\$d} will contain the sum of \texttt{\$a} and \texttt{\$b}.

The requirement is straightforward: we need an opcode. The opcode will explicitly identify the operation to perform. Additionally the opcode will implicitly define the number of operands. Finally, we need some sort of symbolic identifiers for the operands. Keep in mind that the operands will be multidimensional arrays.

%___________________________________________________________________________

\subsubsection*{\phantomsection%
  Interface%
  \addcontentsline{toc}{subsubsection}{Interface}%
  \label{interface}%
}

The Vector Engine and the Vector Engine Manager exposes simple API that consists of the following functions: initialization, finalization, registration of a user-defined operation and execution of a list of bytecodes. Furthermore, the Vector Engine Manager exposes a function to define new arrays.

%___________________________________________________________________________

\subsubsection*{\phantomsection%
  Bridge%
  \addcontentsline{toc}{subsubsection}{Bridge}%
  \label{bridge}%
}

The Bridge is the \textbf{bridge} between the programming interface, e.g. Python/NumPy, and the Vector Engine Manager. The Bridge is the only component that is specifically implemented for the programming interface. In order to add cphVB support to a new language or library, one only has to implement the bridge component. It generates bytecode based on programming interface and sends them to the Vector Engine Manager.

%___________________________________________________________________________

\subsubsection*{\phantomsection%
  Vector Engine Manager%
  \addcontentsline{toc}{subsubsection}{Vector Engine Manager}%
  \label{vector-engine-manager}%
}

Instead of allowing the front-end to communicate directly with the Vector Engine, we introduce a Vector Engine Manager (VEM) into the design. It is the responsibility of the VEM to manage data ownership and distribute bytecode instructions to several Vector Engines. It is also the ideal place to implement code optimization, which will benefit all Vector Engines.

To facilitate late allocation, and early release of resources, the VEM handles instantiation and destruction of arrays. At array creation only the meta data is actually created. Often arrays are created with structured data (e.g. random, constants), with no data at all (e.g. empty), or as a result of calculation. In any case it saves, potentially several, memory copies to delay the actual memory allocation. Typically, array data will exist on the computing device exclusively.

In order to minimize data copying we introduce a data ownership scheme. It keeps track of which components in cphVB that needs to access a given array. The goal is to allow the system to have several copies of the same data while ensuring that they are in synchronization. We base the data ownership scheme on two instructions, \textbf{sync} and \textbf{discard}:%
\begin{description}
\item[{Sync}] \leavevmode 

is issued by the bridge to request read access to a data object. This means that when acknowledging a \textbf{sync} request, the copy existing in shared memory needs to be the most resent copy.
\item[{Discard}] \leavevmode 

is used to signal that the copy in shared memory has been updated and all other copies are now invalid. Normally used by the bridge to upgrading a read access to a write access.
\end{description}

The cphVB components follow the following four rules when implementing the data ownership scheme:\newcounter{listcnt0}
\begin{list}{\arabic{listcnt0}.}
{
\usecounter{listcnt0}
\setlength{\rightmargin}{\leftmargin}
}

\item 

The Bridge will always ask the Vector Engine Manager for access to a given data object. It will send a \textbf{sync} request for read access, followed by a \textbf{release} request for write access. The Bridge will not keep track of ownership itself.
\item 

A Vector Engine can assume that it has write access to all of the output parameters that are referenced in the instructions it receives. Likewise, it can assume read access on all input parameters.
\item 

A Vector Engine is free to manage its own copies of arrays and implement its own scheme to minimize data copying. It just needs to copy modified data back to share memory when receiving a \textbf{sync} instruction and delete all local copies when receiving a \textbf{discard} instruction.
\item 

The Vector Engine Manager keeps track of array ownership for all its children. The owner of an array has full (i.e. write) access. When the parent component of the Vector Engine Manager, normally the Bridge, request access to an array, the Vector Engine Manager will forward the request to the relevant child component. The Vector Engine Manager never accesses the array itself.\end{list}

Additionally, the Vector Engine Manager needs the capability to handle multiple children components. In order to maximize parallelism the Vector Engine Manager can distribute workload and array data between its children components.

%___________________________________________________________________________

\subsubsection*{\phantomsection%
  Vector Engine%
  \addcontentsline{toc}{subsubsection}{Vector Engine}%
  \label{vector-engine}%
}

Though the Vector Engine is the most complex component of cphVB, it has a very simple and a clearly defined role. It has to execute all instructions it receives in a manner that obey the serialization dependencies between instructions. Finally, it has to ensure that the rest of the system has access to the results as governed by the rules of the \textbf{sync}, \textbf{release}, and \textbf{discard} instructions.

%___________________________________________________________________________

\subsection*{\phantomsection%
  Implementation of cphVB%
  \addcontentsline{toc}{subsection}{Implementation of cphVB}%
  \label{implementation-of-cphvb}%
}

In order to demonstrate our cphVB design we have implemented a basic cphVB setup. This concretization of cphVB is by no means exhaustive. The setup is targeting the NumPy library executing on a single machine with multiple CPU-cores. In this section, we will describe the implementation of each component in the cphVB setup – the Bridge, the Vector Engine Manager, and the Vector Engine. The cphVB design rules (Sec. Design) govern the interplay between the components.

%___________________________________________________________________________

\subsubsection*{\phantomsection%
  Bridge%
  \addcontentsline{toc}{subsubsection}{Bridge}%
  \label{id25}%
}

The role of the Bridge is to introduce cphVB into an already existing project. In this specific case NumPy, but could just as well be \texttt{R} or any other language/tool that works primarily on vectorizable operations on large data objects.

It is the responsibility of the Bridge to generate cphVB instructions on basis of the Python program that is being run. The NumPy Bridge is an extension of NumPy version 1.6. It uses hooks to divert function call where the program access cphVB enabled NumPy arrays. The hooks will translate a given function into its corresponding cphVB bytecode when possible. When it is not possible, the hooks will feed the function call back into NumPy and thereby forcing NumPy to handle the function call itself.

The Bridge operates with two address spaces for arrays: the cphVB space and the NumPy space. All arrays starts in the NumPy space as a default. The original NumPy implementation handles these arrays and all operations using them. It is possible to assign an array to the cphVB space explicitly by using an optional cphVB parameter in array creation functions such as \texttt{empty} and \texttt{random}. The cphVB bridge implementation handles these arrays and all operations using them.

In two circumstances, it is possible for an array to transfer from one address space to the other implicitly at runtime.\begin{quotation}%
\begin{quote}
\setcounter{listcnt0}{0}
\begin{list}{\arabic{listcnt0}.}
{
\usecounter{listcnt0}
\setlength{\rightmargin}{\leftmargin}
}

\item 

When an operation accesses an array in the cphVB address space but it is not possible for the bridge to translate the operation into cphVB code. In this case, the bridge will synchronize and move the data to the NumPy address space. For efficiency no data is actually copied instead the bridge uses the \texttt{mremap}\DUfootnotemark{id26}{id30}{†} function to re-map the relevant memory pages.
\item 

When an operations access arrays in different address spaces the Bridge will transfer the arrays in the NumPy space to the cphVB space. Afterwards, the bridge will translate the operation into bytecode that cphVB can execute.\end{list}

\end{quote}
\end{quotation}

In order to detect direct access to arrays in the cphVB address space by the user, the original NumPy implementation, a Python library or any other external source, the bridge protects the memory of arrays that are in the cphVB address space using \texttt{mprotect}\DUfootnotemark{id27}{id31}{‡}. Because of this memory protection, subsequently accesses to the memory will trigger a segmentation fault. The Bridge can then handle this kernel signal by transferring the array to the NumPy address space and cancel the segmentation fault. This technique makes it possible for the Bridge to support all valid Python/NumPy application since it can always fallback to the original NumPy implementation.

In order to gather greatest possible information at runtime, the Bridge will collect a batch of instructions rather than executing one instruction at a time. The Bridge will keep recording instruction until either the application reaches the end of the program or untranslatable NumPy operations forces the Bridge to move an array to the NumPy address space. When this happens, the Bridge will call the Vector Engine Manager to execute all instructions recorded in the batch.

%___________________________________________________________________________

\subsubsection*{\phantomsection%
  Vector Engine Manager%
  \addcontentsline{toc}{subsubsection}{Vector Engine Manager}%
  \label{id28}%
}

The Vector Engine Manager (VEM) in our setup is very simple because it only has to handle one Vector Engine thus all operations go to the same Vector Engine. Still, the VEM creates and deletes arrays based on specification from the Bridge and handles all meta-data associated with arrays.

%___________________________________________________________________________

\subsubsection*{\phantomsection%
  Vector Engine%
  \addcontentsline{toc}{subsubsection}{Vector Engine}%
  \label{id29}%
}

In order to maximize the CPU cache utilization and enables parallel execution the first stage in the VE is to form a set of instructions that enables data blocking. That is, a set of instructions where all instructions can be applied on one data block completely at a time without violating data dependencies. This set of instructions will be referred to as a kernel.

The VE will form the kernel based on the batch of instructions it receives from the VEM. The VE examines each instruction sequentially and keep adding instruction to the kernel until it reaches an instruction that is not \textbf{blockable} with the rest of the kernel. In order to be blockable with the rest of the kernel an instruction must satisfy the following two properties where $A$ is all instructions in the kernel and $N$ is the new instruction.\setcounter{listcnt0}{0}
\begin{list}{\arabic{listcnt0}.}
{
\usecounter{listcnt0}
\setlength{\rightmargin}{\leftmargin}
}

\item 

The input arrays of $N$ and the output array of $A$ do not share any data or represents precisely the same data.
\item 

The output array of $N$ and the input and output arrays of $A$ do not share any data or represents precisely the same data.\end{list}

When the VE has formed a kernel, it is ready for execution. Since all instruction in a kernel supports data blocking the VE can simply assign one block of data to each CPU-core in the system and thus utilizing multiple CPU-cores. In order to maximize the CPU cache utilization the VE may divide the instructions into even more data blocks. The idea is to access data in chunks that fits in the CPU cache. The user, through an environment variable, manually configures the number of data blocks the VE will use.%
\DUfootnotetext{id30}{id26}{†}{
The function mremap() in GNU C library 2.4 and greater.}
\DUfootnotetext{id31}{id27}{‡}{
The function mprotect() in the POSIX.1-2001 standard.}

%___________________________________________________________________________

\subsection*{\phantomsection%
  Performance Study%
  \addcontentsline{toc}{subsection}{Performance Study}%
  \label{performance-study}%
}
\begin{table}
\setlength{\DUtablewidth}{\linewidth}
\begin{longtable*}[c]{|p{0.365\DUtablewidth}|p{0.272\DUtablewidth}|}
\hline

Processor & 

Intel Core i5-2510M \\
\hline

Clock & 

2.3 GHz \\
\hline

Private L1 Data Cache & 

128 KB \\
\hline

Private L2 Data Cache & 

512 KB \\
\hline

Shared L3 Cache & 

3072 KB \\
\hline

Memory Bandwidth & 

21.3 GB/s \\
\hline

Memory & 

4GB DDR3-1333 \\
\hline

Compiler & 

GCC 4.6.3 \\
\hline
\end{longtable*}
\caption{ASUS P31SD. \DUrole{label}{tab:specs}}\end{table}

In order to demonstrate the performance of our initial cphVB implementation and thereby the potential of the cphVB design, we will conduct some performance benchmarks using NumPy\DUfootnotemark{id32}{id34}{§}. We execute the benchmark applications on ASUS P31SD with an Intel Core i5-2410M processor (Table \DUrole{ref}{tab:specs}).

The experiments used the three vector engines: \DUroletitlereference{simple}, \DUroletitlereference{score} and \DUroletitlereference{mcore} and for each execution we calculate the relative speedup of cphVB compared to NumPy. We perform strong scaling experiments, in which the problem size is constant though all the executions. For each experiment, we find the block size that results in best performance and we calculate the result of each experiment using the average of three executions.

The benchmark consists of the following Python/NumPy applications. All are pure Python applications that make use of NumPy and none uses any external libraries.\begin{quotation}%
\begin{quote}
\begin{itemize}

\item 

\textbf{Jacobi Solver} An implementation of an iterative jacobi solver with fixed iterations instead of numerical convergence. (Fig. \DUrole{ref}{benchmark:jacobi}).
\item 

\textbf{kNN} A naive implementation of a k Nearest Neighbor search (Fig. \DUrole{ref}{benchmark:knn}).
\item 

\textbf{Shallow Water} A simulation that simulates a system governed by the shallow water equations. It is a translation of a MATLAB application by Burkardt \cite{Bur10} (Fig. \DUrole{ref}{benchmark:swater}).
\item 

\textbf{Synthetic Stencil} A synthetic stencil simulation the code relies heavily on the slicing operations of NumPy. (Fig. \DUrole{ref}{benchmark:stencil}).
\end{itemize}

\end{quote}
\end{quotation}

%___________________________________________________________________________

\subsubsection*{\phantomsection%
  Discussion%
  \addcontentsline{toc}{subsubsection}{Discussion}%
  \label{discussion}%
}

The jacobi solver shows an efficient utilization of data-blocking to an extent competing with using multiple processors. The \DUroletitlereference{score} engine achieves a 1.42x speedup in comparison to NumPy ($3.98sec$ to $2.8sec$).

On the other hand, our naive implementation of the k Nearest Neighbor search is not an embarrassingly parallel problem. However, it has a time complexity of $O(n^2)$ when the number of elements and the size of the query set is $n$, thus the problem should be scalable. The result of our experiment is also promising – with a performance speedup of of 3.57x ($5.40sec$ to $1.51sec$) even with the two single-core engines and a speed-up of nearly 6.8x ($5.40sec$ to $0.79$)  with the multi-core engine.

The Shallow Water simulation only has a time complexity of $O(n)$ thus it is the most memory intensive application in our benchmark. Still, cphVB manages to achieve a performance speedup of 1.52x ($7.86sec$ to $5.17sec$) due to memory-allocation optimization and 2.98x ($7.86sec$ to $2.63sec$) using the multi-core engine.

Finally, the synthetic stencil has an almost identical performance pattern as the shallow water benchmark the \DUroletitlereference{score} engine does however give slightly better results than the \DUroletitlereference{simple} engine. Score achieves a speedup of 1.6x ($6.60sec$ to $4.09sec$) and the \DUroletitlereference{mcore} engine achieves a speedup of 3.04x ($6.60sec$ to $2.17sec$).

It is promising to observe that even most basic vector engine (\DUroletitlereference{simple}) shows a speedup and in none of our benchmarks a slowdown. This leads to the promising conclusion that the memory optimizations implemented outweigh the cost of using cphVB. Adding the potential of speedup due to data-blocking motivates studying further optimizations in addition to thread-level-parallelization.
The \DUroletitlereference{mcore} engine does provide speedups, the speedup does however not scale with the number of cores. This result is however expected as the benchmarks are memory-intensive and the memory subsystem is therefore the bottleneck and not the number of computational cores available.\begin{figure}[]
\noindent{\includegraphics[width=\columnwidth]{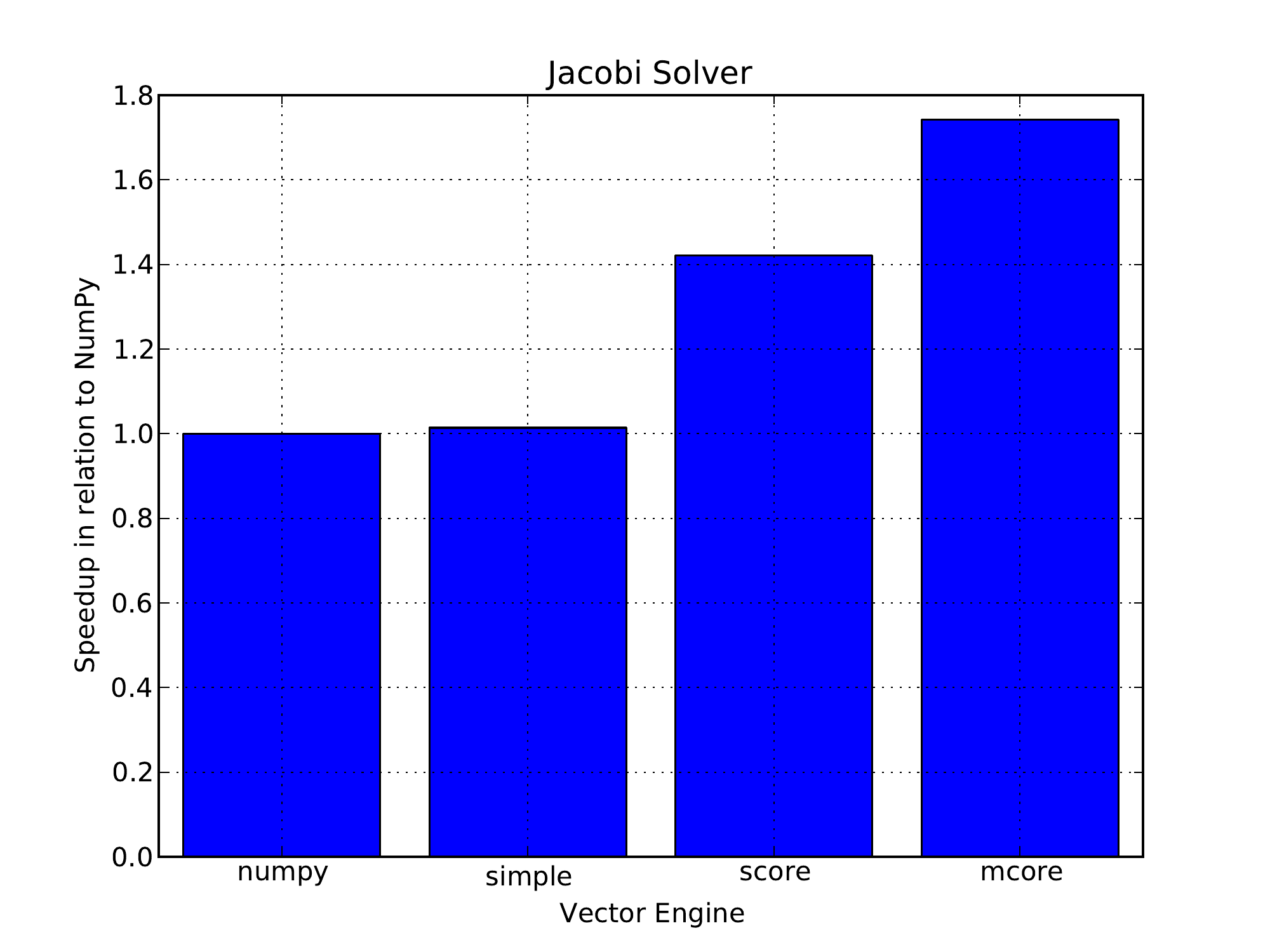}\hfill}
\caption{Relative speedup of the Jacobi Method. The job consists of a vector with $7168x7168$ elements using four iterations. \DUrole{label}{benchmark:jacobi}}
\end{figure}
\begin{figure}[]
\noindent{\includegraphics[width=\columnwidth]{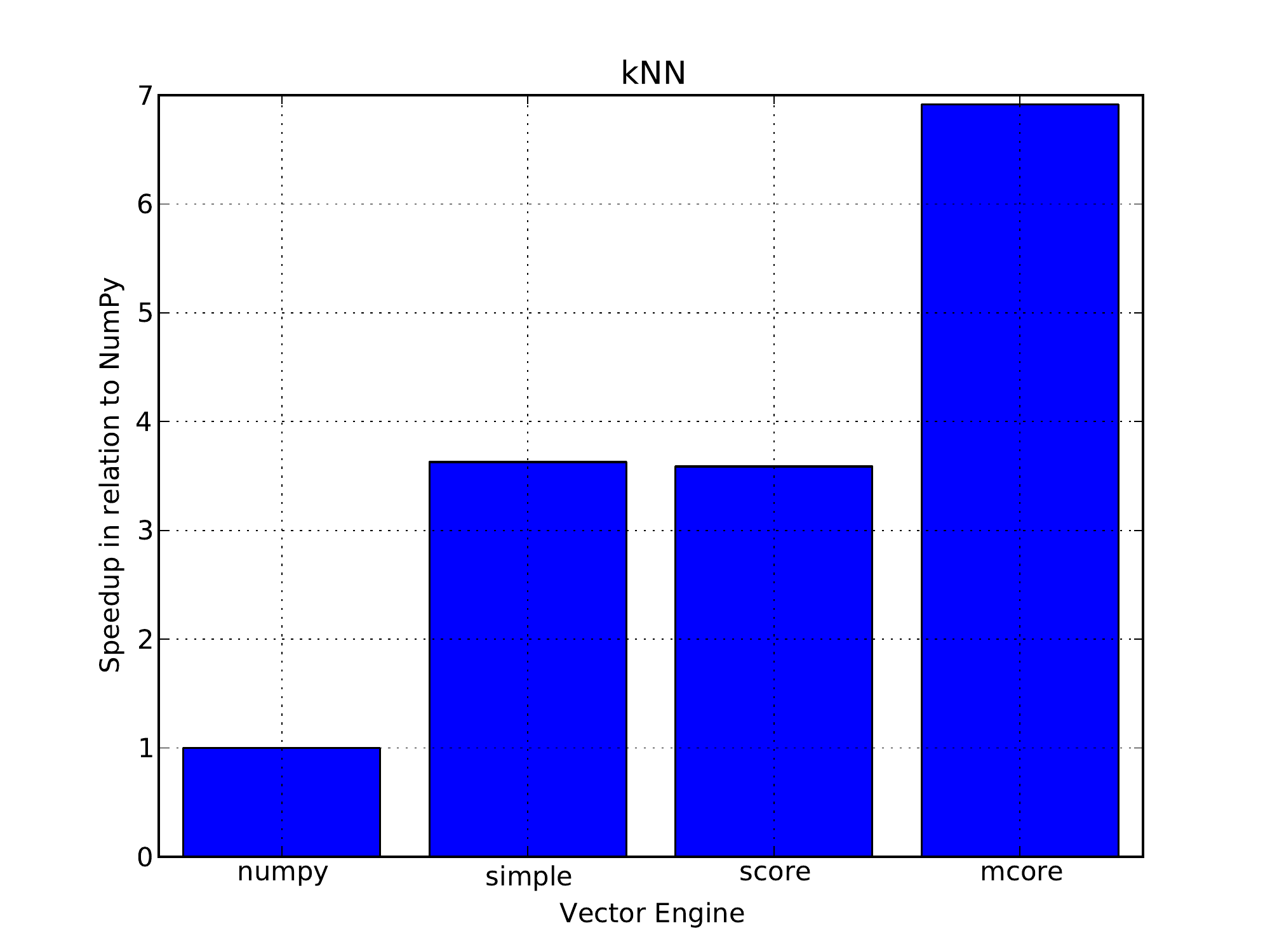}\hfill}
\caption{Relative speedup of the k Nearest Neighbor search. The job consists of 10.000 elements and the query set also consists of 1K elements. \DUrole{label}{benchmark:knn}}
\end{figure}
\begin{figure}[]
\noindent{\includegraphics[width=\columnwidth]{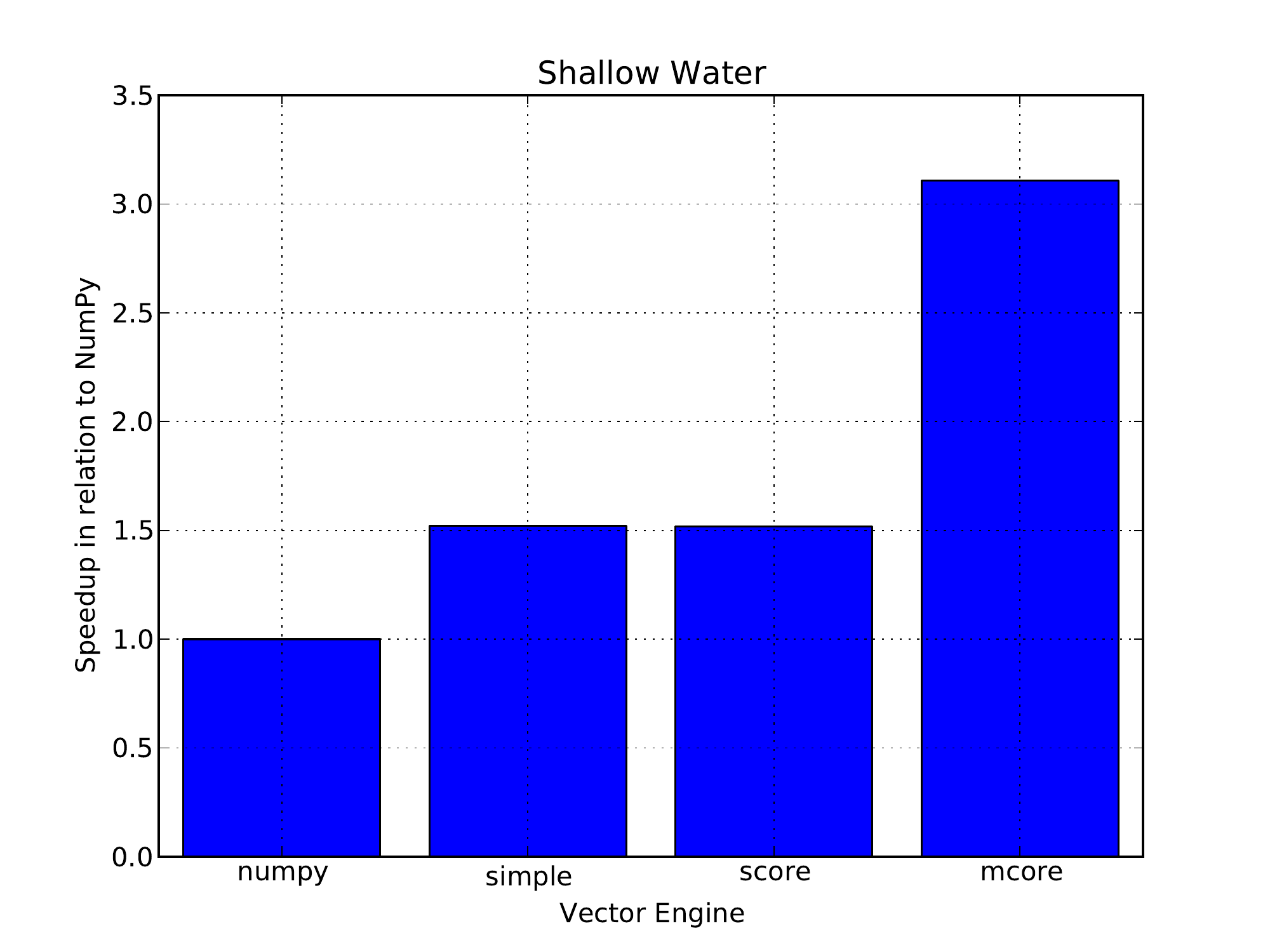}\hfill}
\caption{Relative speedup of the Shallow Water Equation. The job consists of 10.000 grid points that simulate 120 time steps. \DUrole{label}{benchmark:swater}}
\end{figure}
\begin{figure}[]
\noindent{\includegraphics[width=\columnwidth]{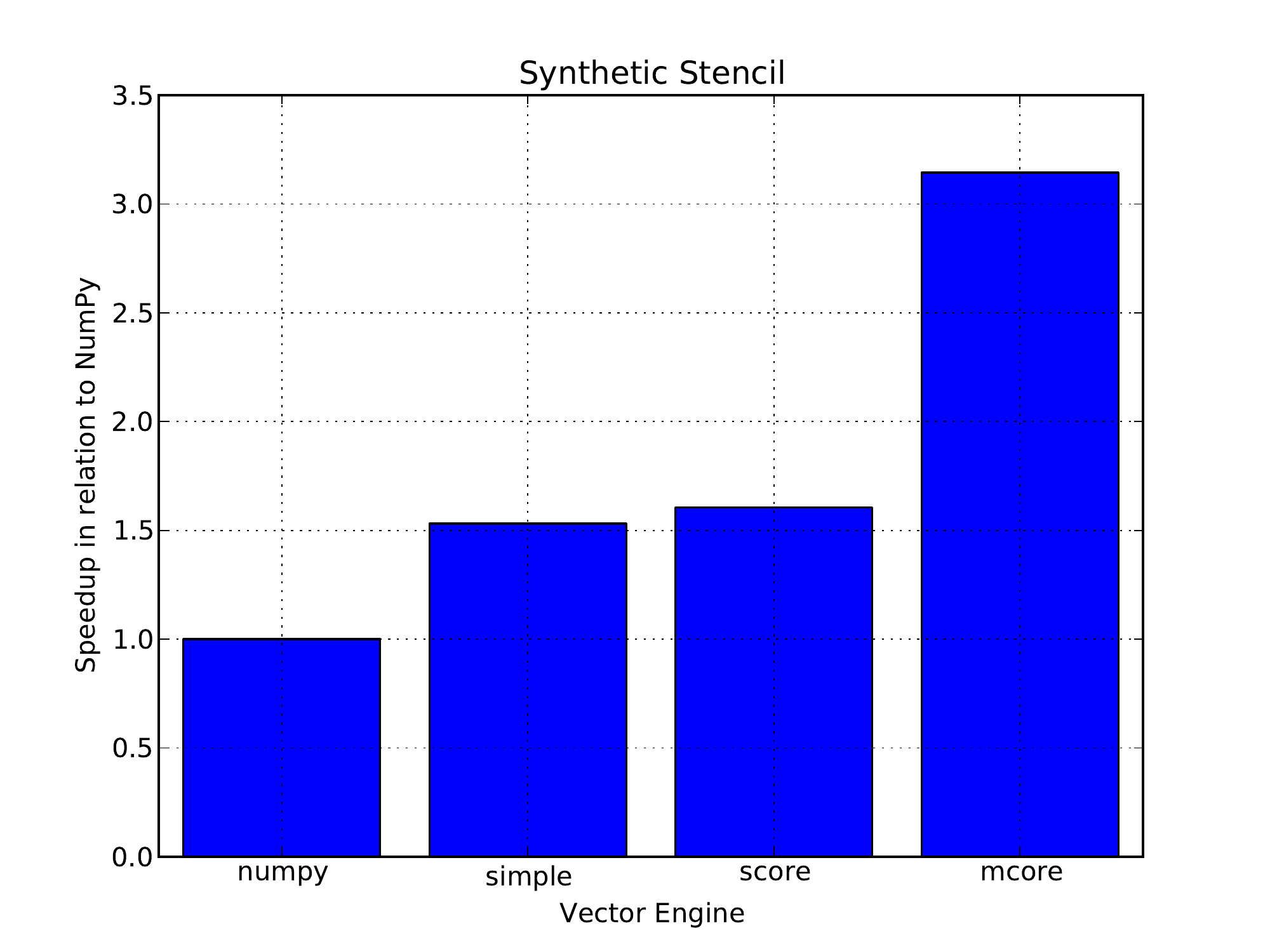}\hfill}
\caption{Relative speedup of the synthetic stencil code. The job consists of vector with $10240x1024$ elements that simulate 10 time steps. \DUrole{label}{benchmark:stencil}}
\end{figure}
\DUfootnotetext{id34}{id32}{§}{
NumPy version 1.6.1.}

%___________________________________________________________________________

\subsection*{\phantomsection%
  Future Work%
  \addcontentsline{toc}{subsection}{Future Work}%
  \label{future-work}%
}

The future goals of cphVB involves improvement in two major areas; expanding support and improving performance. Work has started on a CIL-bridge which will leverage the use of cphVB to every CIL based programming language which among others include: C\#, F\#, Visual C++ and VB.NET. Another project in current progress within the area of support is a C++ bridge providing a library-like interface to cphVB using operator overloading and templates to provide a high-level interface in C++.

To improve both support and performance, work is in progress on a vector engine targeting OpenCL compatible hardware, mainly focusing on using GPU-resources to improve performance. Additionally the support for program execution using distributed memory is on progress. This functionality will be added to cphVB in the form a vector engine manager.

In terms of pure performance enhancement, cphVB will introduce JIT compilation in order to improve memory intensive applications. The current vector engine for multi-cores CPUs uses data blocking to improve cache utilization but as our experiments show then the memory intensive applications still suffer from the von Neumann bottleneck \cite{Bac78}. By JIT compile the instruction kernels, it is possible to improve cache utilization drastically.

%___________________________________________________________________________

\subsection*{\phantomsection%
  Conclusion%
  \addcontentsline{toc}{subsection}{Conclusion}%
  \label{conclusion}%
}

The vector oriented programming model used in cphVB provides a framework for high-performance and high-productivity. It enables the end-user to execute vectorized applications on a broad range of hardware architectures efficiently without any hardware specific knowledge. Furthermore, the cphVB design supports scalable architectures such as clusters and supercomputers. It is even possible to combine architectures in order to exploit hybrid programming where multiple levels of parallelism exist. The authors in \cite{Kri11} demonstrate that combining shared memory and distributed memory parallelism through hybrid programming is essential in order to utilize the Blue Gene/P architecture fully.

In a case study, we demonstrate the design of cphVB by implementing a front-end for Python/NumPy that targets multi-core CPUs in a shared memory environment. The implementation executes vectorized applications in parallel without any user intervention. Thus showing that it is possible to retain the high abstraction level of Python/NumPy while fully utilizing the underlying hardware. Furthermore, the implementation demonstrates scalable performance – a k-nearest neighbor search purely written in Python/NumPy obtains a speedup of more than five compared to a native execution.

Future work will further test the cphVB design model as new front-end technologies and heterogeneous architectures are supported.


\begin{thebibliography}{Kri10}
\bibitem[Kri10]{Kri10}{

M. R. B. Kristensen and B. Vinter, \emph{Numerical Python for Scalable Architectures},
in Fourth Conference on Partitioned Global Address Space Programming Model, PGAS\{’\}10. ACM, 2010. {[}Online{]}. Available: \url{http://distnumpy.googlecode.com/files/kristensen10.pdf}}
\bibitem[Dav04]{Dav04}{

T. David, P. Sidd, and O. Jose, \emph{Accelerator : Using Data Parallelism to Program GPUs for General-Purpose Uses},
October. {[}Online{]}. Available: \url{http://research.microsoft.com/apps/pubs/default.aspx?id=70250}}
\bibitem[New11]{New11}{

C. J. Newburn, B. So, Z. Liu, M. Mccool, A. Ghuloum, S. D. Toit, Z. G. Wang, Z. H. Du, Y. Chen, G. Wu, P. Guo, Z. Liu, and D. Zhang, \emph{Intels Array Building Blocks : A Retargetable , Dynamic Compiler and Embedded Language},
Symposium A Quarterly Journal In Modern Foreign Literatures, pp. 1–12, 2011. {[}Online{]}. Available: \url{http://software.intel.com/en-us/blogs/wordpress/wp-content/uploads/2011/03/ArBB-CGO2011-distr.pdf}}
\bibitem[Klo09]{Klo09}{

A. Kloeckner, N. Pinto, Y. Lee, B. Catanzaro, P. Ivanov, o and A. Fasih, \emph{PyCUDA and PyOpenCL: A Scripting-Based Approach to GPU Run-Time Code Generation},
Brain, vol. 911, no. 4, pp. 1–24, 2009. {[}Online{]}. Available: \url{http://arxiv.org/abs/0911.3456}}
\bibitem[Khr10]{Khr10}{

K. Opencl, W. Group, and A. Munshi, \emph{OpenCL Specification},
ReVision, pp. 1–377, 2010. {[}Online{]}. Available: \url{http://scholar.google.com/scholar?hl=en&btnG=Search&q=intitle:OpenCL+Specification}\#2}
\bibitem[Nvi10]{Nvi10}{

N. Nvidia, \emph{NVIDIA CUDA Programming Guide 2.0},
pp. 1–111, 2010. {[}Online{]}. Available: \url{http://developer.download.nvidia.com/compute/cuda/32}prod/toolkit/docs/CUDACProgrammingGuide.pdf}
\bibitem[Ros10]{Ros10}{

G. V. Rossum and F. L. Drake, \emph{Python Tutorial},
History, vol. 42, no. 4, pp. 1–122, 2010. {[}Online{]}. Available: \url{http://docs.python.org/tutorial/}}
\bibitem[Int08]{Int08}{

Intel, \emph{Intel Math Kernel Library (MKL)},
pp. 2–4, 2008. {[}Online{]}. Available: \url{http://software.intel.com/en-us/articles/intel-mkl/}}
\bibitem[Mat10]{Mat10}{

MATLAB, version 7.10.0 (R2010a).
Natick, Massachusetts: The MathWorks Inc., 2010.}
\bibitem[Rrr11]{Rrr11}{

R Development Core Team, \emph{R: A Language and Environment for Statistical Computing, R Foundation for Statistical Computing},
Vienna, Austria, 2011. {[}Online{]}. Available: \url{http://www.r-project.org}}
\bibitem[Idl00]{Idl00}{

B. A. Stern, \emph{Interactive Data Language},
ASCE, 2000.}
\bibitem[Oct97]{Oct97}{

J. W. Eaton, \emph{GNU Octave},
History, vol. 103, no. February, pp. 1–356, 1997. {[}Online{]}. Available: \url{http://www.octave.org}}
\bibitem[Oli07]{Oli07}{

T. E. Oliphant, \emph{Python for Scientific Computing},
Computing in Science Engineering, vol. 9, no. 3, pp. 10–20, 2007. {[}Online{]}. Available: \url{http://ieeexplore.ieee.org/lpdocs/epic03/wrapper.htm?arnumber=4160250}}
\bibitem[Gar10]{Gar10}{

R. Garg and J. N. Amaral, \emph{Compiling Python to a hybrid execution environment},
Computing, pp. 19–30, 2010. {[}Online{]}. Available: \url{http://portal.acm.org/citation.cfm?id=1735688.1735695}}
\bibitem[Pas05]{Pas05}{

R. V. D. Pas, \emph{An Introduction Into OpenMP},
ACM SIGARCH Computer Architecture News, vol. 34, no. 5, pp. 1–82, 2005. {[}Online{]}. Available: \url{http://portal.acm.org/citation.cfm?id=1168898}}
\bibitem[Cat09]{Cat09}{

B. Catanzaro, S. Kamil, Y. Lee, K. Asanov'i, J. Demmel, c K. Keutzer, J. Shalf, K. Yelick, and O. Fox, \emph{SEJITS: Getting Productivity and Performance With Selective Embedded JIT Specialization},
in Proc of 1st Workshop Programmable Models for Emerging Architecture PMEA, no. UCB/EECS-2010-23, EECS Department, University of California, Berkeley. Citeseer, 2009. {[}Online{]}. Available: \url{http://www.eecs.berkeley.edu/Pubs/TechRpts/2010/EECS-2010-23.html}}
\bibitem[And08]{And08}{

R. Andersen and B. Vinter, \emph{The Scientific Byte Code Virtual Machine},
in Proceedings of the 2008 International Conference on Grid Computing \& Applications, GCA2008 : Las Vegas, Nevada, USA, July 14-17, 2008. CSREA Press., 2008, pp. 175–181. {[}Online{]}. Available: \url{http://dk.migrid.org/public/doc/published_papers/sbc.pdf}}
\bibitem[Apl00]{Apl00}{

“why apl?”
{[}Online{]}. Available: \url{http://www.sigapl.org/whyapl.htm}}
\bibitem[Sci02]{Sci02}{

R. Pozo and B. Miller, \emph{SciMark 2.0},
2002. {[}Online{]}. Available: \url{http://math.nist.gov/scimark2/}}
\bibitem[Bur10]{Bur10}{

J. Burkardt, \emph{Shallow Water Equations},
2010. {[}Online{]}. Available: \url{http://people.sc.fsu.edu/~jburkardt/m_src/shallow_water_2d/}}
\bibitem[Bac78]{Bac78}{

J. Backus, \emph{Can Programming be Liberated from the von Neumann Style?: A Functional Style and its Algebra of Programs},
Communications of the ACM, vol. 16, no. 8, pp. 613–641, 1978.}
\bibitem[Kri11]{Kri11}{

M. Kristensen, H. Happe, and B. Vinter, \emph{Hybrid Parallel Programming for Blue Gene/P},
Scalable Computing: Practice and Experience, vol. 12, no. 2, pp. 265–274, 2011.}
\end{thebibliography}
\end{document}